\documentclass[aps,prb,twocolumn,floatfix,amsmath,amssymb,showpacs]{revtex4}
\usepackage{graphicx}
\usepackage{dcolumn}
\usepackage{bm}
\usepackage{epsfig,psfrag,amsmath,amssymb,float}
\input{epsf}

\begin{document}
\title{Evolution of pairing from weak to strong coupling on a honeycomb lattice}

\author{Shi-Quan Su$^{a,b}$, Ka-Ming Tam$^{c}$ and Hai-Qing Lin$^a$}

\affiliation{$^a$ Department of Physics and Institute of Theoretical Physics,
The Chinese University of Hong Kong, Hong Kong, China\\
$^b$ Department of Physics and Astronomy, Louisiana State University, Baton Rouge, Louisiana 70803\\
$^c$ Department of Physics and Astronomy, University of Waterloo, Waterloo, Ontario, N2L-3G1, Canada}

\begin{abstract}
We study the evolution of the pairing from weak to strong coupling on a honeycomb
lattice by Quantum Monte Carlo. We show numerical evidence of the BCS-BEC crossover as the coupling strength
increases on a honeycomb lattice with small fermi surface by measuring a wide range
of observables: double occupancy, spin susceptibility, local pair correlation, and
kinetic energy. Although at low energy, the model sustains Dirac fermions, we do not
find significant qualitative difference in the BCS-BEC crossover as compared to
those with an extended Fermi surface, except at weak coupling, BCS regime. 
\end{abstract}
\date{today}

\maketitle

\section{INTRODUCTION}

It has long been known that the pairing formed from an attractive coupling has
a smooth crossover between the weak coupling and the strong
coupling\cite{Eagles,Leggett,Nozieres}. In the weak coupling limit, singlet pairs are
formed around the fermi surface, according to the BCS theory. In the strong
coupling limit, local bound pairs can be formed, and these "preformed pairs"
condense as the temperature is further lowered where the Bose-Einstein
condensation(BEC) occurs. The interest on this crossover has been revitalized
\cite{psgapRev1,psgapRev2,SymmDimPsgap,DMFTspingap,PairScAtt,Kyung,Garg}, mainly due to
the quest of understanding the pseudogap phase in the high temperature
superconductors.

Recently, condensed matter systems sustain on fermions with linear dispersion, typical examples are honeycomb lattice models and 
nodal fermions for $d$-wave superconductors, have generated huge surge of intensive studies. These models possess substantial differences
from models with extended Fermi surface such as models on square lattice. In particular, it has been suggested that the quantum phase transition
(QPT) between the metallic phase and the degenerate charge density wave/pairing phase at half-filling in the attractive Hubbard model (AHM) on
honeycomb
lattice is related to its BCS-BEC crossover away from half-filling \cite{MFHexCO}. This certainly does not happen on the square lattice,
in which the flat Fermi surface at half-filling renders the Umklapp scattering becoming the dominant channel, its BCS-BEC crossover is
not related to any QPT through tuning the attractive coupling \cite{SquKTT}. In the honeycomb lattice, the density of state is zero at 
half-filling, therefore any instability from the band structure is weakened, and strong coupling is needed to induce ordering. It can be
shown that all the short range interactions are irrelevant. In order to tackle the strong coupling problem, besides breaking the symmetry by mean
field ansatz, we choose Quantum Monte Carlo method in this work to study the BCS-BEC crossover in the honeycomb lattice.

Various studies \cite{DMFTspingap,PairSpinGap,BSSAtt,KePePair} have been devoted to the BCS-BEC crossover of the AHM on a square lattice. The
objective of this work is to study how do the linear dispersion, and the aforementioned QPT at half-filling affect the BCS-BEC 
crossover of the slightly doped system.

Our main finding can be summarized as follow. At the weak coupling, BCS-like regime, pseudogap phenomena are observed, however we 
expect that it is mainly due to the band structure of honeycomb lattice, rather than the bound pair formation. At the intermediate coupling, 
crossover regime, we can identify two temperature scales, the high temperature one where the performed pair formed with associated pseudogap 
phenomena; and the low temperature one where the system enters the pairing phase. At strong coupling, BEC-like regime, the electrons form pairs at 
high temperature and condense as hard core bosons at low temperature. However, we do not find distinctive feature compares to the square lattice, 
except at the weak coupling regime where the band structure dominates the quasi-particle dispersion. Further interpretations of the QMC results 
are next presented by applying the mean field (MF) approximation to lattice models and continuum model for fermions with linear dispersion.

\section{MODEL and METHOD}
The AHM in honeycomb lattice reads
\begin{equation}
H =-t\sum_{<i,j>,\sigma} c^+_ {i\sigma}c_{j\sigma} -U\sum_i
n_{i\uparrow}n_{i\downarrow}-\mu\sum_{i\sigma}n_{i\sigma},
\label{eq:hamitl2}
\end{equation}
where $c_{i\sigma}(c^+_{i\sigma})$ annihilates (creates) a particle with spin
$\sigma$ at site $i$, $\left\langle i,j \right\rangle$ denotes the
nearest-neighbor lattice sites $i$ and $j$, $t$ is the hopping matrix element,
$U$ is the on-site attractive interaction, and $ \mu $ is the chemical
potential. In the following we set $t=1$ as the energy scale of the system, all
the observable are in units of $t$. The bare electronic ($U=0$ limit)
dispersion is given by $\epsilon_{\textbf{k}} = \pm\sqrt{3+2cos(\sqrt{3}k_{y}) +
4cos(\sqrt{3}k_{y}/2)cos(3k_{x}/2)}$, and the band width $W$ is 6. At half
filling this is linear around the Fermi points. Keeping only the low energy
excitations, in the first quantized form the wave function follows the 2D Weyl
equation for massless chiral Dirac fermions, $v_{F} \widehat{\sigma} \cdot
\nabla \Psi(\mathbf{r})=E\Psi(\mathbf{r})$, where
$\widehat{\sigma}=(\sigma_{x},\sigma_{y})$ are the Pauli matrices and
$v_{F}=3/2$ is the Fermi velocity. This description in term of Dirac fermions is not exact away from half filling. Nevertheless,
the linear dispersion can be a good approximation below the van Hove
singularities at filling $n=1\pm1/4$. For this reason, we choose $n=0.88$ for
our calculations using determinant quantum Monte Carlo(DQMC)
\cite{Blankenbecler}. 

\begin{figure} [bth]
\includegraphics[width=7.2cm,height=6.0cm]{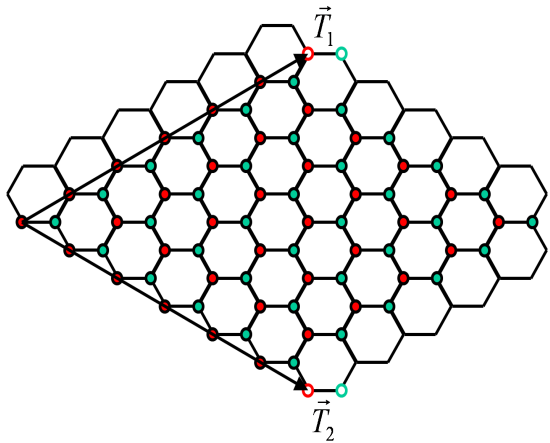}
\caption{ Sketch of a $72$ sites honeycomb lattice. The red and green solid circles are the 
lattice points in the honeycomb lattice. The
red solid circles also represent the underlying triangular lattice. $\vec{T}_{1}$ and $\vec{T}_{2}$ are the real space 
translational vectors.}
\label{hex72}
\end{figure}

The DQMC \cite{Blankenbecler,Loh} is a Hamiltonian based approach. The Hamiltonian $H$ in the partition function $Z=Tr\exp(-\beta H)$
is expressed in the real space via the Trotter decomposition and Hubbard-Stratonovich (HS) transformation. The only systematic error
is from discretizing the imaginary-time $\beta$ into $M$ slices of $\Delta\tau=\beta/M$ in the Trotter decomposition. The HS transformation 
replaces the on-site interactions in the attractive Hubbard model by HS fields coupled to the charge. The summation over the HS fields is 
treated by Monte Carlo procedure. The calculations are proceeded on a $N=72$ sites honeycomb lattice, the actual lattice for the simulation is 
shown in the Fig. \ref{hex72}. Since the attractive Hubbard coupling does not have minus-sign problem, a wide range of temperatures and couplings 
can be studied.

\section{QMC RESULTS}
One of the clear signals indicating the formation of bound pairs at strong coupling is the formation of spin gap. At weak coupling, we expect 
fermion quasi-particle character to remain at high temperature, for which the spin susceptibility increases as the temperature is lowered. On the 
other hand, the strong coupling limit is manifested by the decrease of the spin susceptibility as the temperature is lowered, due to the formation 
of the gap which leads to the reduction in the spectral function at low frequency. 

\begin{figure} [bth]
\includegraphics[width=9.0cm,height=6.0cm]{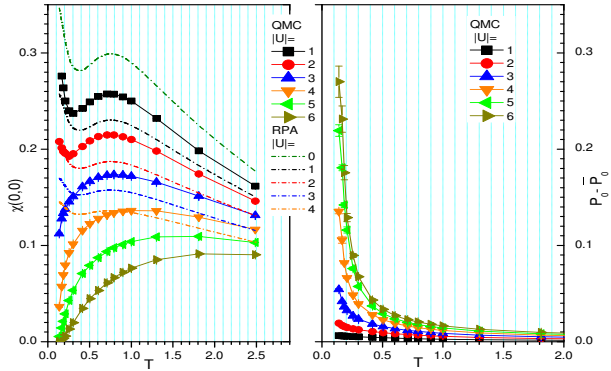}
\caption{Uniform spin susceptibility $\chi(0,0)$ (left), and pairing
correlation function $ P_0-\bar{P}_0 $ (right) as a function of temperature for
a range of interaction strength at $n=0.88$. }
\label{ChiP}
\end{figure}

We first show the spin susceptibility $\chi(\textbf{q},\omega)$ 
at frequency $\omega=0$, and momentum $\textbf{q}=(0,0)$ in Fig. \ref{ChiP}, where we also show the spin susceptibility from RPA calculation for 
comparison. $\chi(0,0)$ is suppressed for all couplings, as can be inferred simply from the RPA formulation, where 
$\chi_{RPA}(0,0)=\chi_{0}(0,0)/(1+U\chi_{0}(0,0))$. At weak coupling $\chi(0,0)$ increases as the temperature is lowered as expected for a fermion
quasi-particle description, however it bends downward before it goes upward again as the temperature is lowered further. This two peak structure of
$\chi(0,0)$ associated with the formation of the pseudogap has been found in the dynamical mean field theory study\cite{DMFTspingap}. However, in 
the honeycomb lattice, the apparent pseudogap phenomena indicated by this structure of $\chi(0,0)$ already exist in the weak coupling regime, 
below the strong coupling regime where the "preformed pair" phenomena occur. Therefore, we believe that it is derived from the particular 
dispersion relation of honeycomb lattice, where the density of state is small around the doped Fermi surface.

On the other hand, in the strong coupling regime, $\chi(0,0)$ vanishes quickly as the bound pairs are formed and spin gap equals
the binding energy needed to break the pair. In the weak coupling regime, the QMC results behave similarly as compared to the
RPA results. When the interaction is increased to around $W/2$, the QMC results evolve in the opposite direction as compared
to the RPA results and drop sharply at low temperature, whereas the RPA results at low temperature limit do not change qualitatively when $U$
increases. This signals that the system enters the phase in which electrons form bound pairs, and the spin excitations start to
be gapped \cite{PairSpinGap}. The pairing phase cannot be reached by summing the ladder diagrams within the RPA. For
strong coupling ($U \approx W$), the suppression of $\chi(0,0)$ becomes smooth and appears at high temperature. This effect
reflects the fact that the bound pairs are already formed at high temperature \cite{BSSAtt}. The temperature where deviations appear
between the QMC results and the RPA results is an indication of the formation of local singlet pair, which can be interpreted as the
energy scale where the fermion quasi-particle description is not valid for any lower temperature.

We then probe the pairing directly by considering the pairing
correlation function for local pairing, $P_0 =\frac{1}{N^2}
\sum_{l,i} \langle
c^{\dagger}_{i+l,\uparrow}c^{\dagger}_{i+l,\downarrow}
c_{i,\downarrow}c_{i,\uparrow} +h.c. \rangle$. The only instability
is pairing, in this incommensurate doped case (rules out CDW
order). We expect $P_0$ to increase as temperature is lowered for
all coupling strengths. One of the most representative
characteristics of local pairs is that they are distributed uniformly
in space and condense around zero momentum as bosons when
temperature is lowered. This is manifested by the rapid increase of
local pair correlation as shown in Fig. \ref{ChiP}, note that the single
particle contribution $\bar{P}_0$ has been subtracted from $P_0$ to
emphasize the vertex contribution of pairing \cite{White}. The
condensation of the bosonic local pairs for the strong coupling case
($U=5,6$) can be observed from the rapid increase of $P_0-\bar{P_0}$
with the decrease of temperature. In contrast, the pairs formed
around the quasi-particle Fermi surface in the weak coupling regime
only bring a slight increase in $P_0-\bar{P_0}$.

\begin{figure} [bth]
\includegraphics[width=9.0cm,height=6.0cm]{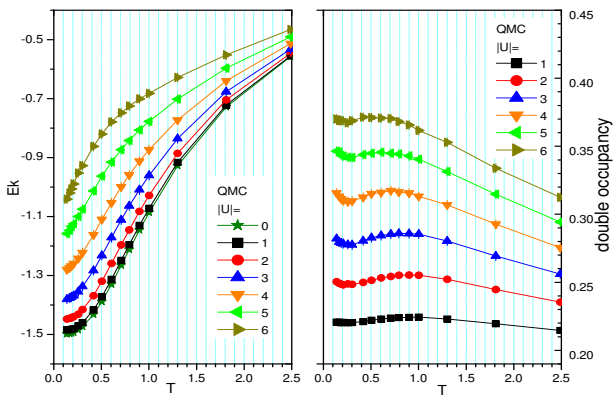}
\caption{The kinetic energy(left), and double occupancy(right)
as a function of temperature for a range of interaction strength at $n =0.88$.}
\label{EkN}
\end{figure}

We show the kinetic energy, $E_k=(-t/N)\sum_{\langle i, j\rangle,\sigma} \langle c^+_{i,\sigma}c_{j,\sigma} \rangle$ in Fig. \ref{EkN}. In the
weak coupling regime, its temperature dependence is similar to the free fermion case. When we increase the interaction to the crossover regime
($U\approx3-4$), qualitative change already happened in the high temperature, where the gain in the kinetic energy is much slower than the free
fermion case. Moving into the strong coupling regime, fermions begin to form bound pairs at high temperature and only lose little kinetic energy.
When temperature further decreases, the local pairs in the system condense and hence $E_k$ drops sharply.
\cite{BSSAtt}.

A good indicator to measure the local pair formation in the BEC
state is the double occupancy $\langle n_{\uparrow} n_{\downarrow}
\rangle$, see Fig. \ref{EkN}. We find that $\langle n_{\uparrow} n_{\downarrow} \rangle
$ increases as the temperatures decrease. However, it
reaches a local maximum at certain temperature. This can be
understood as the change of the kinetic energy which destabilizes
the double occupancy. This behavior of $\langle n_{\uparrow}
n_{\downarrow} \rangle $ are in accord with the fact that the local
maximum coincides with the temperature where the kinetic energy
drops most sharply. At very low temperature, the bosonic on-site
pairs begin to dominate, $\langle n_{\uparrow} n_{\downarrow}
\rangle $ increases again and should saturate at $n/2$ for strong
couplings.

\begin{figure}
\includegraphics[width=9.0cm]{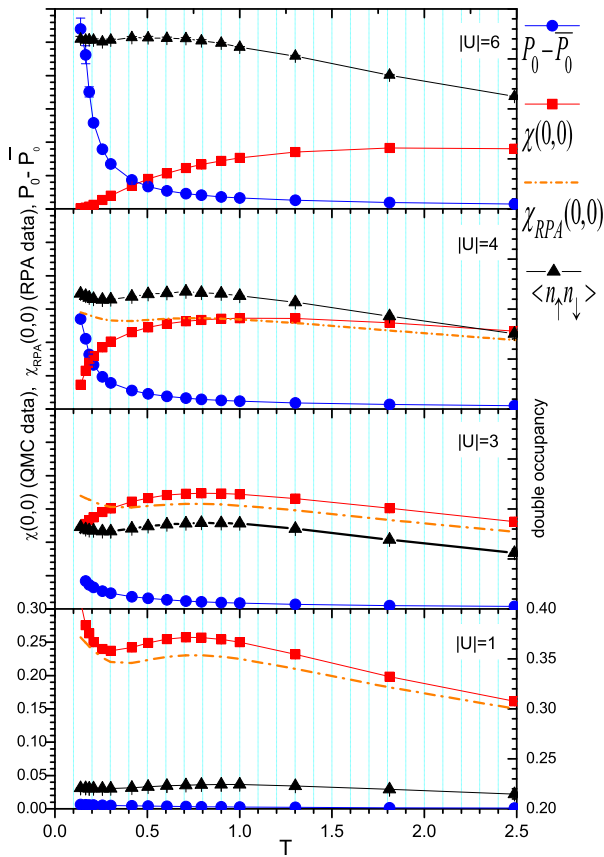}
\caption{\label{3} Double occupancy $\langle n_{\uparrow}n_{\downarrow}
\rangle$, uniform spin susceptibility $\chi(0,0)$, and pairing correlation
$P_{0}-\overline{P_{0}}$ as a function of temperature for different coupling
strength at $\langle n \rangle = 0.88$ filling. The magenta shadow regions are
used to mark the energy scale.  }
\label{All}
\end{figure}

After elaborating the evidence of BCS-BEC crossover, we put those observables from DQMC together and identify the temperature scales for
different $U$. In Fig. \ref{All}, we show the results represented for weak ($U=1$),
intermediate ($U=3, 4$), and strong ($U=6$) couplings.

At $U=1$, $\chi(0,0)$ QMC result does not deviate from the RPA result. The
local pair correlation does not develop, and $\langle n_{\uparrow}n_{\downarrow} \rangle$ is small even at low temperature, which shows that the
pairing correlation is weak. The critical temperature, $T_{c}$, for the Kosterlitz-Thouless transition into the pairing phase is below the
lowest temperatures we studied.

At $U=3$, $\chi(0,0)$ local pair correlations begin to increase at $T_c\approx0.2$. At almost the same temperature the
QMC result begins to deviate from the RPA result. These imply the developments in both spin and pairing correlations. The system shows BCS-like
pairing effect from the instability of the fermi surface. However, there is no true phase coherence at any finite temperature as that in the BCS
theory. Nevertheless, at this coupling strength, the pairing is still rather weak, due to the small density of state around the Fermi energy.

At $U=4$ the system displays two temperature scales. The first one is $T^{*}$ at high temperature around $T\approx0.8$, this could be associated
with the pseudogap phase. At this temperature, $\chi(0,0)$ from QMC result reaches its maximum and begins to deviate from the RPA result. In
addition $\langle n_{\uparrow} n_{\downarrow} \rangle $ also reaches the first plateau at high temperature. These signal that electrons bound
pairs start to develop, spin gap is formed and the quasi-particle description is broken below this temperature. We estimate the critical
temperature for the condensation of bound pairs, $T_c \approx 0.3$. Below this temperature, the local pair correlation $P_0-\bar{P}_0$ grows
quickly and $\chi(0,0)$ drops sharply; $\langle n_{\uparrow}n_{\downarrow}\rangle $ reaches its low temperature maximum and saturates.

At $U=6$, the system is at the strong coupling limit, where $U$ reaches the band width $W$, there is only one temperature scale in
the system, $T_c\approx0.5$, within the temperature range we studied. $\chi(0,0)$ reaches its maximum at very high temperature
and decreases smoothly, which suggests that pair formation begins at a very high temperature, above the temperature range we studied.
Below $T_c$, $P_0-\bar{P}_0$ increase quickly, and $\langle n_{\uparrow}n_{\downarrow}\rangle$ tends to $n/2$ at zero
temperature. These suggest that the bound pairs undergo a Kosterlitz-Thouless transition, which manifests a BEC-like scenario.

From the above numerically exact DQMC data, we show clearly that
there is a qualitative change from weak to strong coupling at finite temperature.
This should correspond to the true BCS-BEC crossover at zero temperature.
However, we find that the results for the honeycomb lattice have no drastic qualitative difference as compared to
that of the square lattice \cite{BSSAtt}. Certainly, the band structure alters the
quantitative values of the coupling for the crossover. However, the BCS-BEC crossover on a doped honeycomb lattice models exists at $U\approx3-4$
where the linear dispersion approximation for the free fermions is not valid.

\section{DISCUSSIONS AND CONCLUSIONS}
With the progress of the techniques of optical lattices and the fabrication of
single layer graphene, the BCS-BEC crossover on a honeycomb lattice and Dirac
fermions is not only an important problem itself, but also has broad
experimental and theoretical interests with other topics under intensive
studies. 

The atom-atom interaction in ultracold fermionic atoms in a optical trap
can be tuned by magnetic field Feshbach resonance. The honeycomb lattice can
possibly be realized by optical trap \cite{hexOLRamaSpectroscopy}. This may provide a direct way
to study experimentally the BCS-BEC crossover problem with linear dispersion. \cite{hexOLRamaSpectroscopy}.
In addition, the superconducting phase of graphene via the attraction from phonons and plasmons has been discussed 
recently\cite{Uchoa1,Uchoa2,Kopnin,Bergman,Uchoa3}.
Although it is unlikely to generate strong attraction from phonon coupling in graphene, our results suggest that even at weak coupling regime,
non-trivial temperature dependence of spin susceptibility may occur in the superconducting phase from local Holstein phonon coupling.

In conclusion, we have presented extensive results from DQMC which confirm the BCS-BEC crossover for the doped (n=0.88) AHM on a honeycomb lattice.
In contrast to the systems with extended fermi surface, there is an enhancement of pseudogap property revealed from the double peak structure in
the spin susceptibility at weak coupling due to the peculiar density of state of honeycomb lattice. Apart from this, the BCS-BEC crossover does
not show prominent difference between square lattice and honeycomb lattice for the parameters and system size we study.


\section{ACKNOWLEDGEMENTS}
The authors thank Z.-B. Huang, S.-J. Gu, P. McClarty, and Y.-Z. You
for useful discussions; and thank F. Ng at the ITSC of CUHK, where
the numerical work presented in this paper was accomplished. This
work is supported by HKSAR RGC Project CUHK 401806. The research at the University of Waterloo was funded by the
Canada Research Chair Program (M. Gingras, Tier 1) and
the Shared Hierarchical Academic Research Computing
Network.

\end{document}